\documentclass[10pt,letterpaper,twocolumn]{article} 

\usepackage{ol2}
\usepackage[draft]{hyperref}
\usepackage{amsmath}

\begin{document}

\twocolumn[ 

\title{Non-Hermitian Multimode Interference}


\author{Stefano Longhi}
\address{Dipartimento di Fisica, Politecnico di Milano and Istituto di Fotonica e Nanotecnologie del Consiglio Nazionale delle Ricerche, Piazza L. da Vinci 32, I-20133 Milano, Italy (stefano.longhi@polimi.it)\\ and
IFISC (UIB-CSIC), Instituto de Fisica Interdisciplinar y Sistemas Complejos, E-07122 Palma de Mallorca, Spain}
\author{Liang Feng}
\address{Department of Materials Science and Engineering, University of Pennsylvania, Philadelphia, Pennsylvania 19104, USA}

\begin{abstract}
Multi-mode interference (MMI) and self-imaging are important phenomena of diffractive wave optics with major applications
in optical signal processing, beam shaping and optical sensing. Such phenomena generally arise from interference of 
 normal modes in lossless dielectric guiding structures, however the impact of spatially-inhomogeneous optical gain and loss, 
 which break mode orthogonality and symmetries, has been overlooked. Here we consider MMI in non-Hermitian optical 
 systems, either graded index or coupled optical waveguide structures with optical gain and loss, and reveal distinctive features, such as 
 the absence of mirror images and strong-sensitivity of self-imaging to perturbations, making MMI in non-Hermitian waveguides of interest in optical sensing.  
\end{abstract}

 ] 

{\it Introduction.} Multi-mode interference (MMI) and self-imaging effects are well-known phenomena of diffractive wave optics \cite{r1,r2,r3,r4} which find major applications
in different fields ranging from optical imaging and sensing \cite{r4,r5,r6,r7,r8,r8bis} to beam shaping \cite{r9,r10}, laser design \cite{r11,r12}, optical signal processing \cite{r3,r13,r14}, and quantum information \cite{r16}. In most of previous studies, interference arises among normal modes of lossless dielectric waveguides or fibers, which are orthogonal modes like the eigenmodes of a Hermitian system. In the past decade, parity-time ($\mathcal{PT}$) and non-Hermitian optics have emerged as flourishing research fields, where judiciously-tailored spatial regions of optical gain and loss in dielectric guiding systems deeply change the flow of light, with the demonstration of unprecedented optical functionalities (see e.g. \cite{r18,r19,r20,r21,r22,r23,r24} for recent reviews). Non-orthogonality of modes in a non-Hermitian optical system is responsible for a wide variety of intriguing effects, such as the appearance of exceptional points \cite{r21,r22}, unidirectional scattering \cite{r25,r26,r27,r28}, excess noise and enhanced sensitivity to perturbations \cite{r29,r30,r31,r32,r33,r34,r35}. Self-imaging effects in non-Hermitian optical systems have been so far limited to consider periodic wave fields, corresponding to the non-Hermitian extension of the Talbot effect \cite{r36,r37}, however MMI among non-orthogonal modes in guiding non-Hermitian structures and distinctive features of non-Hermitian versus Hermitian MMI have been so far overlooked.\par 
In this Letter we extend MMI and self-imaging effects to the non-Hermitian realm by considering graded index or coupled optical waveguide structures with balanced optical gain and loss, synthesized to show real and commensurate propagation constants for all guided modes. As compared to the their Hermitian counterpart, distinctive features arise from non-orthogonality of guided modes, such as the absence of mirror images and a strong-sensitivity of self-imaging to perturbations, which could be of major advantage for sensing applications.
 \par
  \begin{figure}[htbp]
\centerline{\includegraphics[width=8.6cm]{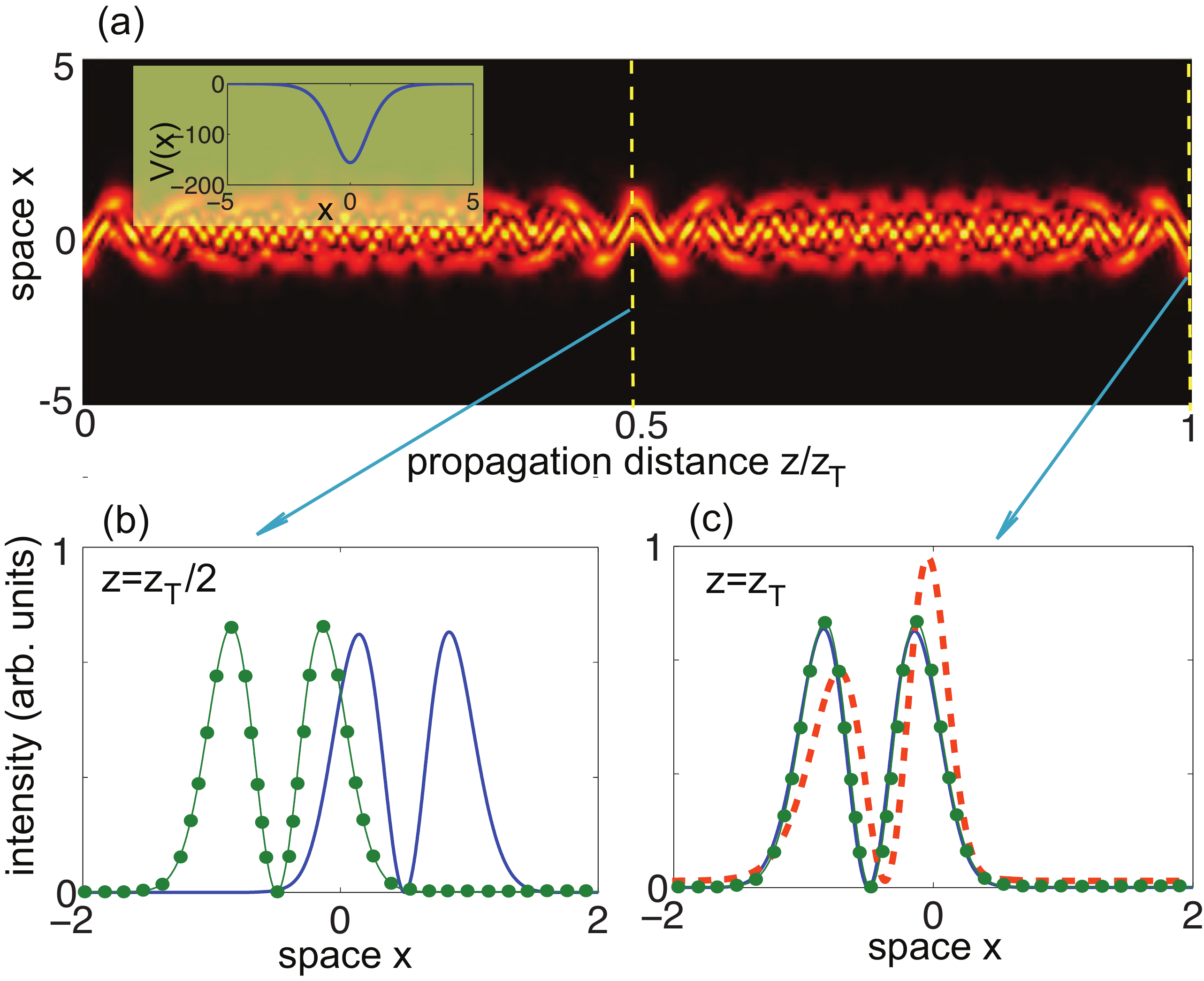}} \caption{ \small
(Color online) Self-imaging in an Hermitian P\"oschl-Teller graded-index waveguide sustaining $N=\nu=12$ modes. (a) Numerically-computed evolution of the field amplitude $|\psi(x,z)|$ on a pseudo color map. The initial field distribution is the two-humped profile $\psi(x,0)=(x+d) \exp[-(x+d)^2/w^2]$ with $d=w=0.5$. The inset in (a) shows the waveguide profile $V(x)$. (b,c) Intensity profile $|\psi(x,z)|^2$ (solid curves) at the mirror-image plane $z=z_T/2$ (panel (b)) and at the self-imaging plane $z=z_T$ (panel (c)), with $z_T=2 \pi$. The solid-dotted curves shows the intensity profile of the input field at $z=0$. Note that in (c) the solid-dotted and solid curves are almost overlapped. The thick dashed curve in (c) shows breakdown of self-imaging for a slightly modified value of $\nu$ to the non-integer value $\nu=12.05$.}
\end{figure} 
{\it Self-imaging in non-Hermitian graded-index waveguides.} Self-imaging effects arising from MMI in graded-index waveguides or fibers is observed when the propagation constants of the excited guided modes are commensurable \cite{r3}. In the weak-guiding, scalar and paraxial approximations, light propagation along the optical $Z-$ axis in a graded-index slab waveguide with transverse spatial direction $X$ is governed by the paraxial wave equation \cite{r38}
\begin{equation}
\frac{i}{k} \frac{\partial \psi}{\partial Z}= -\frac{1}{2 n_s k^2} \frac{\partial^2 \psi}{\partial X^2}+\left\{n_s -n(X) \right\} \psi
\end{equation}
for the electric field amplitude $\psi(X,Z)$, where $k=2 \pi / \lambda$ is the photon wave number (in vacuum), $n_s$ is the substrate (cladding) refractive index, and $n(X) \simeq n_s$ the spatially-varying refractive index profile of core region. Indicating by $\Delta n_0$ the characteristic index change between cladding and core regions of the guide, it is worth introducting the characteristic longitudinal and transverse spatial lengths as $L_{\parallel} = \lambda / (2 \pi \Delta n_0)$, $L_{\perp}=\lambda/(2 \pi \sqrt{2 n_s \Delta n_0})$. In dimensionless spatial variables $z=Z/L_{\parallel}$,  $x=X/L_{\perp}$ Eq.(1)
reads
\begin{equation}
i \frac{\partial \psi}{\partial z}= - \frac{\partial^2 \psi}{\partial x^2}+V(x) \psi  \equiv H \psi(x,z)
\end{equation}
where $V(x)  \equiv (n_s-n(x)) / \Delta n_0$ is the normalized optical potential. Let us indicate by $n(x)$ the refractive index profile of a multimode lossless waveguide sustaining $N$ guided modes $u_l(x)$ with propagation constants $E_l$ ($l=0,1,2,...,N-1$) showing self-imaging effects, so that $H$ is self-adjoint, $n(x)$, $V(x)$ and $E_l$ are real numbers, and  $H u_l(x)=E_l u_l(x)$. An example is provided by a graded-index
P\"oschl-Teller potential
\begin{equation}
V(x)=-\frac{\nu(\nu+1)}{\cosh^2 x}
\end{equation}
($\nu$ integer number), which sustains exactly $N=\nu$ bound modes with propagation constants $E_n=-(\nu-n)^2$ ($n=0,1,2,..,\nu-1$). Owing to the symmetry $V(-x)=V(x)$, the guided modes show either odd or even symmetry after spatial inversion, namely $u_n(-x)=(-1)^n u_n(x)$. The quadratic form of propagation constants on mode index $n$ ensures exact self-imaging of any initial field distribution $\psi(x,0)$, which does not excite scattering modes, after a propagation distance $z_T=2 \pi$, i.e. $\psi(x,z_T)=\psi(x,0)$. Likewise, the symmetry properties of eigenmodes under spatial inversion implies that at $z=z_T/2=\pi$ a mirror image of the initial distribution is observed, i.e. $\psi(x,z_T/2)= \pm \psi(-x,0)$. Such imaging effects are explained as a result of MMI \cite{r3}. In fact, an initial field distribution $\psi(x,0)=\sum_{n=0}^{N-1} c_l u_n(x)$, obtained from an arbitrary superposition of guided modes with complex amplitudes $c_l$, propagates along the waveguide according to
\begin{equation}
\psi(x,z)=\sum_{n=0}^{N-1} c_l u_n(x) \exp(-i E_n z).
\end{equation}
For $z=z_T= 2 \pi$, clearly $E_n z_T$ is an integer multiple than $2 \pi$ for any mode number $n$, and thus from Eq.(4) one has $\psi(x,z_T)=\psi(x,0)$ (self-imaging). On the other hand, for $z=z_T/2$ one has $E_n z_T/2=(-1)^{(n+\nu)}$, and thus
\begin{eqnarray}
\psi(x,z_T/2) & = & (-1)^{\nu} \sum_{n, \; even} c_n u_n(x)-(-1)^{\nu}\sum_{n, \; odd} c_n u_n(x) \nonumber \\
& = & (-1)^{\nu} \sum_n c_n u_n(-x),
\end{eqnarray}
 i.e. $\psi(x,z_T/2)= (-1)^{\nu} \psi(-x,0)$ (mirror-imaging). Note that in writing the last equation we used the symmetry property $u_n(-x)=(-1)^n u_n(-x)$ of eigenmodes.
An example of self-imaging and mirror-imaging for a Hermitian graded-index P\"oschl-Teller waveguide is shown in Fig.1.\\ 
Here we wish to extend imaging effects to non-Hermitian multimode graded-index waveguides, where balanced gain and loss regions are added to the guiding refractive index profile. The resulting Hamiltonian $H_1$ is therefore non-Hermitian. For a given lossless graded-index waveguide with refractive index profile $n(x)$ showing self-imaging effects, one can synthesize isospectral waveguides, described by a non-Hermitian Hamiltonian $H_1$ with complex refractive index $n_1(x)$, by means of supersymmetric \cite{r37,r39} or imaginary spatial displacement \cite{r40,r41} methods. For construction, the non-Hermitian waveguide $H_1$ sustains $N$ modes with the same (real) propagation constants $E_n$ as the Hermitian waveguide $H$, and thus self-imaging at the distance $z=z_T$ is still observed. However, the eigenmodes cease to be orthogonal and the odd/even symmetry under spatial inversion is not conserved. This implies that the mirror-image at $z=z_T/2$ is not anymore observed. To exemplify such a result, let us consider a $\mathcal{PT}$-symmetric extension of the  graded-index P\"oschl-Teller waveguide (3), obtained by an imaginary spatial shift \cite{r40}, i.e. let us consider the complex index profile $n_1(x)=n(x+i \delta)$ and corresponding optical potential
\begin{equation}
V_1(x)=V(x+i \delta)=-\frac{\nu(\nu+1)}{\cosh^2 (x+i \delta)}
\end{equation}
 \begin{figure}[htbp]
\centerline{\includegraphics[width=8.6cm]{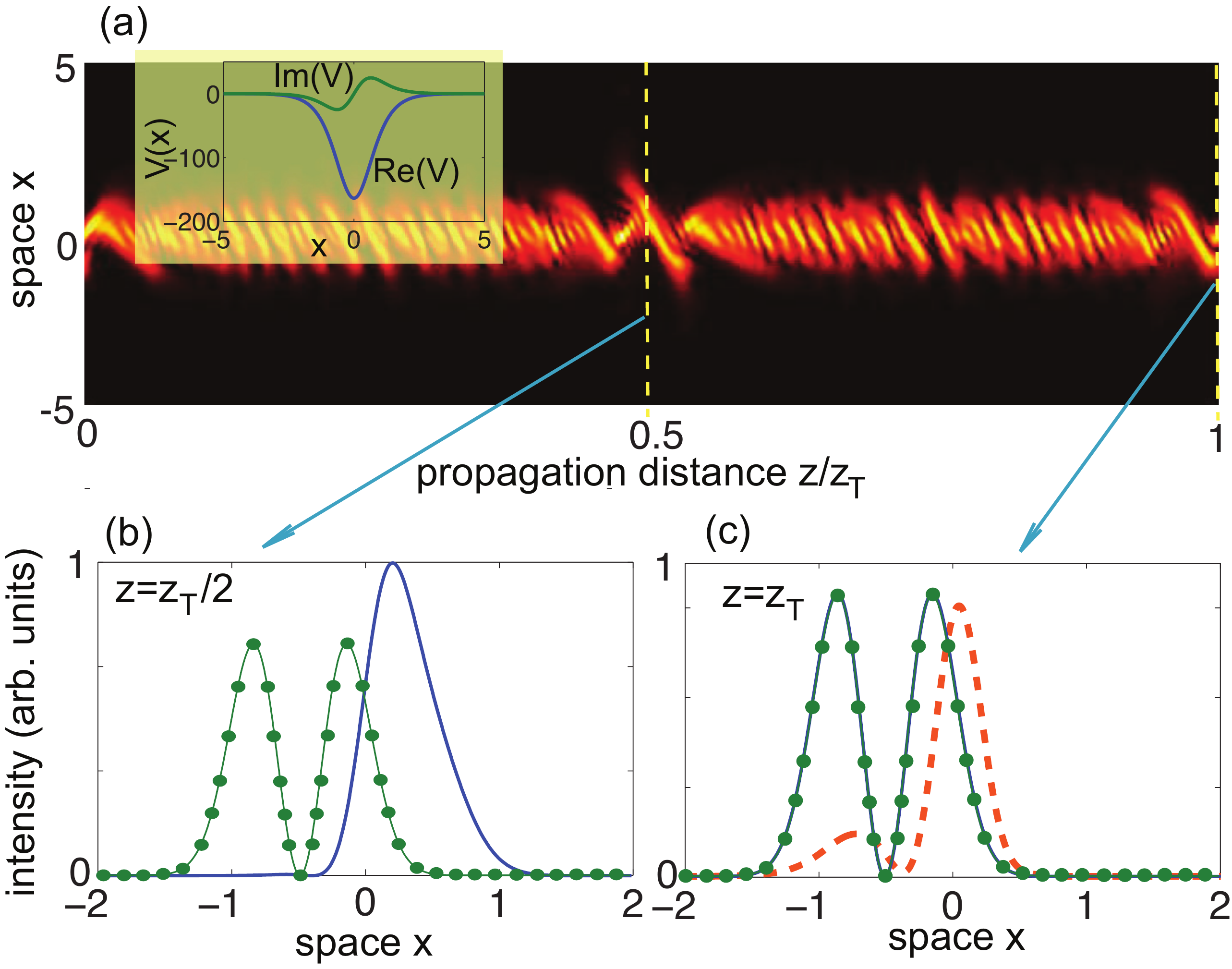}} \caption{ \small
(Color online) Same as Fig.1, but for a $\mathcal{PT}$-symmetric P\"oschl-Teller graded-index waveguide with $\delta=0.2$. For the sake of clearness, in (a) the amplitude $|\psi(x,z)|$ is normalized, at each plane $z$, to $\sqrt{\int dx |\psi(x,z)|^2}$. The inset in (a) shows both real and imaginary parts of $V$.}
\end{figure} 
where $0 \leq \delta < \pi/2$. The Hamiltonians $H_1$ and $H$ are clearly isospectral, and the (right) eigenmodes $v_l(x)$ of $H_1$ are simply obtained from those of $H$ by an imaginary space shift, i.e. 
$v_l(x)=u_l(x+i \delta)$.
Note that, since $V_1(-x)=V_1^*(x)$, the Hamiltonian $H_1$ is $\mathcal{PT}$-symmetric in the unbroken $\mathcal{PT}$ phase and its eigenvectors satisfy the condition $v_l(-x)=(-1)^l v_l^*(x)$.
While self-imaging is observed at the distance $z=z_T=2 \pi$, at the half distance $z=z_T/2$ the mirror-image is not anymore observed rather generally, since in this case one has $\psi(x,z_T/2)=(-1)^{\nu} \psi(-x-2i \delta,0)$. An example of self-imaging and breakdown of mirror-imaging in the $\mathcal{PT}$-symmetric P\"oschl-Teller waveguide is shown in Fig.2. The non-orthogonality of modes is responsible for an enhanced sensitivity of the system to perturbations, and is measured by the Petermann factor \cite{r33,r35,r41} 
\begin{equation}
K_l= \frac{ \langle v_l  (x)|v_l(x) \rangle \langle v^{\dag}_l(x) | v^{\dag}_l(x) \rangle} {|\langle v_l^{\dag}(x) | v_l(x) \rangle |^2}=\frac{ \langle v_l  (x)|v_l(x) \rangle^2}{|\langle v_l^*(x) | v_l(x) \rangle |^2}
\end{equation}
  \begin{figure}[htbp]
\centerline{\includegraphics[width=8.6cm]{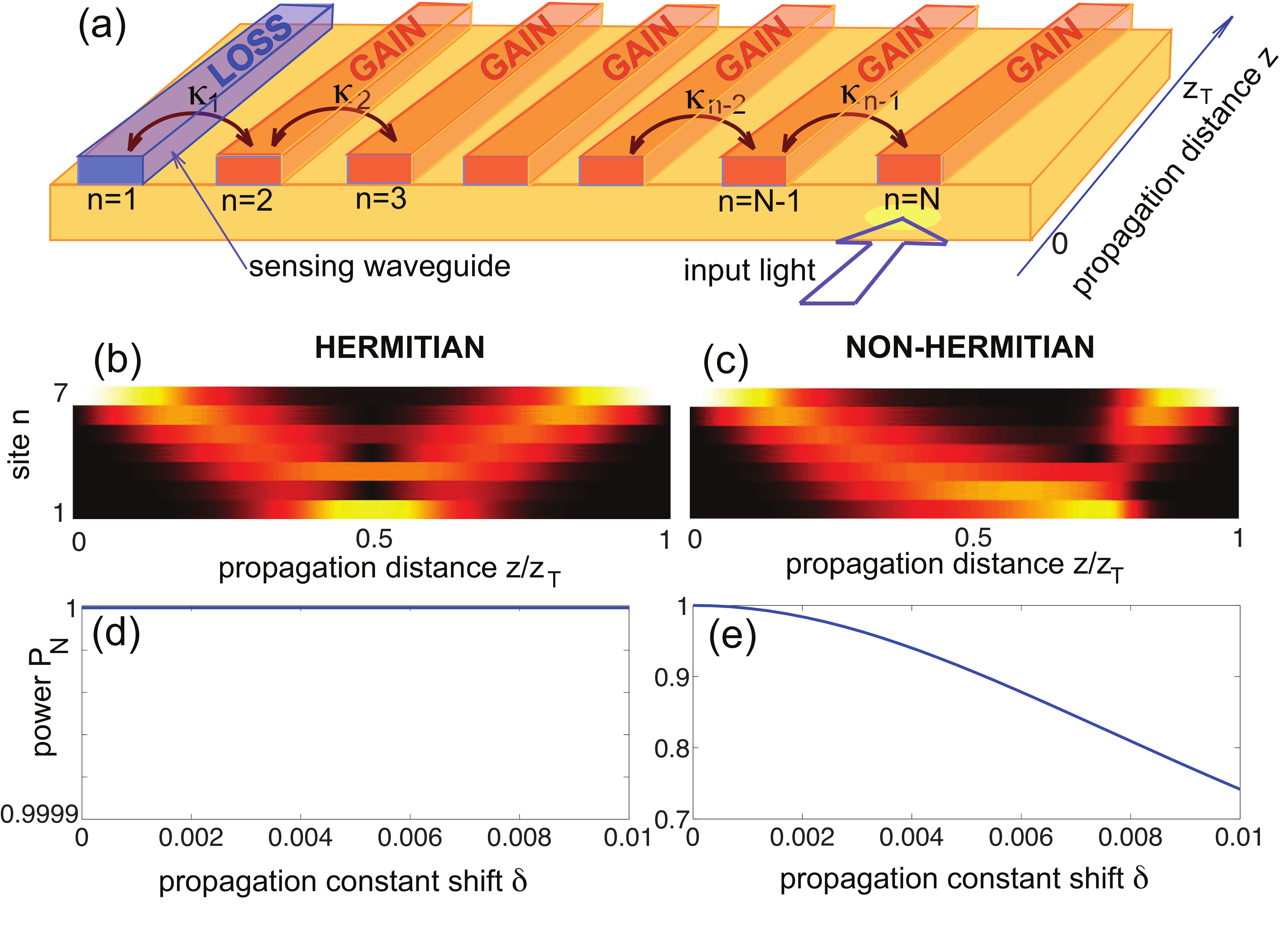}} \caption{ \small
(Color online) (a) Schematic of an array of $N$ waveguides with inhomogeneous coupling constants and with globally balanced gain and loss that realizes self-imaging at the distance $z_T$. The loss parameter in the left edge waveguide is $\gamma$, whereas the gain parameter in the other $(N-1)$ waveguides is $g= \gamma/(N-1)$. For $N=7$, isospectral Hermitian and non-Hermitian waveguide lattices with equally-spaced supermodes, corresponding to the eigenvalues  $E_l=E_0+l \Delta$, ($l=0,1,2,..,7$) with $E_0=-1.5$ and $\Delta=0.5$, are  obtained for coupling constants $\kappa_1=1$, $\kappa_2=0.8660$, $\kappa_3=0.8018$, $\kappa_4=0.7237$, $\kappa_5=0.6155$, $\kappa_6=0.4523$ with gain parameter $g=0$ (Hermitian lattice), and coupling constants  $\kappa_1=2.2361$, $\kappa_2=1.2247$, $\kappa_3=0.9636$, $\kappa_4=0.8092$, $\kappa_5=0.6629$, $\kappa_6=0.4767$ with gain parameter $g=\Delta=0.5$ (non-Hermitian lattice). (c,d) Snapshots of normalized amplitude distribution $|c_n(z)| / \sqrt{\sum_n |c_n(z)|^2}$ along the normalized propagation distance $z/z_T$ for right-edge waveguide input excitation in the Hermitian [panel(b)] and non-Hermitian [panel (c)] waveguide arrays. The self-imaging distance is $z_T= 2 \pi/ \Delta= 4 \pi$. All quantities (i.e. spatial length $z$, coupling constants and gain/loss parameters) are normalized to the coupling $\kappa_1=1$ of the Hermitian lattice. (d,e) Numerically-computed behavior of fractional power $P_N$ versus propagation constant shift $\delta$ for Hermitian and non-Hermitian arrays.}
\end{figure} 
where $v^{\dag}_l(x)=v^*_l(x)$ are the left eigenvectors of $H$. The Petermann factor $K_l$, originally introduced to describe excess quantum noise and excess linewidth in lasers with unstable resonators, provides rather generally a measure of the non-orthogonality degree of the mode of a non-Hermitian system, diverging when the mode becomes self-orthogonal (exceptional point).
In our case  $K_l$ equals one for $\delta=0$ and diverges to infinity as $\delta \rightarrow \pi/2$ ($\mathcal{PT}$ symmetry breaking point). As an example, the bold dashed curves in Figs.1(c) and 2(c) show the numerically-computed self-image, at plane $z=z_T$, for a perturbed refractive index change $\Delta n_0$ of the graded-index waveguide, corresponding to the slightly varied value $\nu=12.05$. Clearly, in the non-Hermitian waveguide [Fig.2(c)] breakdown of self-imaging is more pronounced, even though the Petermann factor is modestly larger than one (for the fundamental mode one has $K_0 \simeq 2.53$). 
\par
{\it MMI and self-imaging in non-Hermitian waveguide lattices.} 
Non-Hermitian multimode graded-index structures, while being a focus of several recent theoretical studies addressing interesting effects as supersymmetry and scatteringless optical potentials \cite{
r28,r39,r40,r40b}, require special tailoring of both real and imaginary parts of the refractive index, and remain so far challenging for practical implementation, with few experiments reported in the microwave spectral region \cite{r41b}.
On the other hand, coupled waveguides or resonators with optical gain and loss offer a more accessible and controllable platform for harnessing non-Hermitian effects \cite{r18,r19,r20,r21,r22,r23,r29,r30,r35}. MMI among array supermodes and self-imaging effects in Hermitian waveguide lattices have been demonstrated in previous experiments (see e.g.\cite{r42}), and applications to chemical sensing have been reported \cite{r8bis}. A main open question is whether self-imaging can occur in non-Hermitian waveguide arrays, and whether non-Hermitian effects could be harnessed in applications such as optical sensing based on MMI. Here we propose a non-Hermitian waveguide lattice setup which shows self-imaging effects in which the imaging condition is hugely sensitive to perturbations as compared to its Hermitian counterpart. The non-Hermitian waveguide lattice is shown in Fig.3(a). It comprises a set of $N$ coupled waveguides with the same propagation constants and inhomogeneous couplings $\kappa_l$, in which the edge waveguide $n=1$ is lossy with a loss parameter $\gamma$, whereas the other $(N-1)$ waveguides experience optical gain with the same gain parameter $g$. We also assume overall balanced gain-loss distribution, i.e. $\gamma=(N-1)g$. Waveguide lattices with precise control of coupling constants can be  manufactured with current semiconductor technology, where selective gain and loss can be accomplished; see for instance \cite{referee}.
Light propagation in the waveguide array is described by coupled-mode equations  for the modal field amplitude in the various waveguides \cite{r8bis,r36,r42}
\begin{equation}
i \frac{dc_n}{dz}= -\kappa_n c_{n+1}-\kappa_{n-1} c_{n-1}-i \gamma_n c_n \equiv H c_n
\end{equation} 
 ($n=1,2,...,N$), with $\kappa_0=\kappa_{N+1}=0$, $\gamma_1=\gamma=(N-1)g$ and $\gamma_n=-g$ for $2 \leq n \leq N$. Like in the graded-index waveguide problem considered above, we wish to synthesize a waveguide lattice such that all propagation constants $E_l$ of array supermodes (i.e. the $N$ eigenvalues of the matrix $H$) are real and commensurate, thus ensuring self-imaging. In particular, we 
 consider here the case where $E_l$ are equally spaced by $\Delta$, i.e. $E_l=E_0+ l \Delta$ ($l=1,2,...,N$).  Clearly, for $N=2$ the problem is trivial since the array reduces to the well-known $\mathcal{PT}$-symmetric optical coupler, which shows self-imaging in the unbroken $\mathcal{PT}$ phase. For $N \geq 3$, the structure is no longer $\mathcal{PT}$ symmetric, but more generally non-Hermitian.  For a given loss rate $\gamma$, the coupling constants $\kappa_l$ that realize an equally-space ladder spectrum can be synthesized by means of inverse spectral methods of Jacobi matrices \cite{r43,r44}. In caption of Fig.3 we provide two examples of isospectral $N=7$ waveguide lattices with supermode spacing $\Delta=0.5$, one Hermitian ($g=0$) and the other non-Hermitian ($g=\Delta$), which have been synthesized by numerically implementing the algorithm described in Ref.\cite{r44}. The two waveguide arrays realize a self-image of any input discrete optical pattern at the output plane $z_T=2 \pi / \Delta$. Examples of self-images are shown in Figs.3(b) and (c). 
 However, in the non-Hermitian array the propagation pattern and self-imaging condition are much more sensitive to perturbations than in the Hermitian array owing to the large values of the Petermann factor arising from mode non-orthogonality. For example, the Petermann factors $K_l$ of supermodes for the non-Hermitian lattice of Fig.3 vary in the range $23-180$. Such large values of $K_l$ suggest an enhanced sensitivity to perturbations \cite{r33,r35}, which could be of interest in MMI-based sensing applications. For example, let us consider a  perturbation to $H$ corresponding to a propagation constant shift $\delta$ of the mode in the lossy waveguide $n=1$. Such a perturbation describes, for example, the case where the lossy waveguide $n=1$ shows a different refractive index change $(dn/dT)$ with temperature than the other waveguides, and the array is used to monitor temperature changes. Likewise, the edge waveguide can act as a chemical transducer element whose effective index changes as the concentration of some chemical compounds varies \cite{r8bis}. The impact of $\delta$ on breakdown of MMI self-imaging can be detected by measuring the fractional light power $P_{N}$ trapped in the right-edge waveguide $n=N$, normalized to the entire optical power in all waveguides at the output plane, i.e. 
 \begin{equation}
 P_N=\frac{|c_N(z_T)|^2}{ \sum_n |c_n(z_T)|^2}
 \end{equation}
when the array is excited, at input plane $z=0$, by the right-edge waveguide $n=N$ [Fig.3(a)]. Figures 3(d,e) show the numerically-computed behavior of $P_N$ versus $\delta$ for Hermitian ($g=0$) and non-Hermitian ($g=\Delta$) arrays, respectively. While in the Hermitian lattice the fractional power $P_N$ is independent of $\delta$ in the considered range, a measurable change of $P_N$ is clearly observed in the non-Hermitian lattice. 
 \par
{\it Conclusions.}
Multimode interference of guided modes in integrated optical structures is of major interest in several applications ranging from beam shaping to optical sensing, signal processing and quantum information. Interesting effects of MMI that are usually exploited in such applications are self-imaging and mirror-imaging phenomena, which are observed when the propagation constants of the excited guided modes are commensurate. In most of current studies of MMI and related applications, light propagation is described within Hermitian models, i.e. neglecting optical loss and gain in the system.  Here we have considered MMI and self-imaging effects in non-Hermitian systems, such as graded-index waveguides and waveguide arrays with inhomogeneous gain and loss distributions. We synthesized suitable structures that realize self-imaging in spite of gain and loss in the system. However, non-orthogonality of modes has a great impact on mode interference. In particular, it has been shown that mirror-imaging is generally prevented in non-Hermitian graded-index structures, and that the interference pattern is much more sensitive to perturbations of system parameters. Our results suggest that the recently developed field of non-Hermitian photonics \cite{r18,r19,r20,r21,r22,r23,r24} could provide interesting and insightful guidelines in the design of MMI-based integrated photonic devices.\\ 
\\
The authors declare no conflicts of interest.\\
\\
L.F. acknowledges support from Army Research Office (W911NF-17-1-0400).\\

\end{document}